\documentclass[journal=ancac3,manuscript=article]{achemso}
\usepackage{graphicx}
\usepackage{epstopdf}
\usepackage{epsfig}
\usepackage{amsmath}
\usepackage{color}
\usepackage{caption}
\DeclareGraphicsExtensions{.pdf,.eps,.png,.jpg,.mps}

\author{Jiang-Bin Wu$^{1}$}
\author{Zhi-Xin Hu$^{2}$}
\author{Xin Zhang$^{1}$}
\author{Wen-Peng Han$^{1}$}
\author{Yan Lu$^{1}$}
\author{Wei Shi$^{1}$}
\author{Xiao-Fen Qiao$^{1}$}
\author{Mari Iji$\ddot{a}$s$^{3}$}
\author{Silvia Milana$^{3}$}
\author{Wei Ji$^{2}$}
\email{wji@ruc.edu.cn}
\author{Andrea C. Ferrari$^{3}$}
\author{Ping-Heng Tan$^{1}$}
\email{phtan@semi.ac.cn}
\affiliation{$^{1}$State Key Laboratory of Superlattices and Microstructures, Institute of Semiconductors, Chinese Academy of Sciences, Beijing 100083, China
\\$^{2}$Department of Physics, Renmin University of China, Beijing 100872, China\\$^{3}$  Cambridge Graphene Centre, University of Cambridge,Cambridge CB3 0FA, UK}

\title{Interface Coupling in Twisted Multilayer Graphene by Resonant Raman Spectroscopy of Layer Breathing Modes}

\begin{document}

\begin{abstract}
Raman spectroscopy is the prime non-destructive characterization tool for graphene and related layered materials. The shear (C) and layer breathing modes (LBMs) are due to relative motions of the planes, either perpendicular or parallel to their normal. This allows one to directly probe the interlayer interactions in multilayer samples. Graphene and other two-dimensional (2d) crystals can be combined to form various hybrids and heterostructures, creating materials on demand with properties determined by the interlayer interaction. This is the case even for a single material, where multilayer stacks with different relative orientations have different optical and electronic properties. In twisted multilayer graphene samples there is a significant enhancement of the C modes due to resonance with new optically allowed electronic transitions, determined by the relative orientation of the layers. Here we show that this applies also to the LBMs, that can be now directly measured at room temperature. We find that twisting has a small effect on LBMs, quite different from the case of the C modes. This implies that the periodicity mismatch between two twisted layers mostly affects shear interactions. Our work shows that Raman spectroscopy is an ideal tool to uncover the interface coupling of 2d hybrids and heterostructures.

\end{abstract}
\subsection{Table of Contents Graphic}
\begin{figure*}
\centerline{\includegraphics[width=140mm,clip]{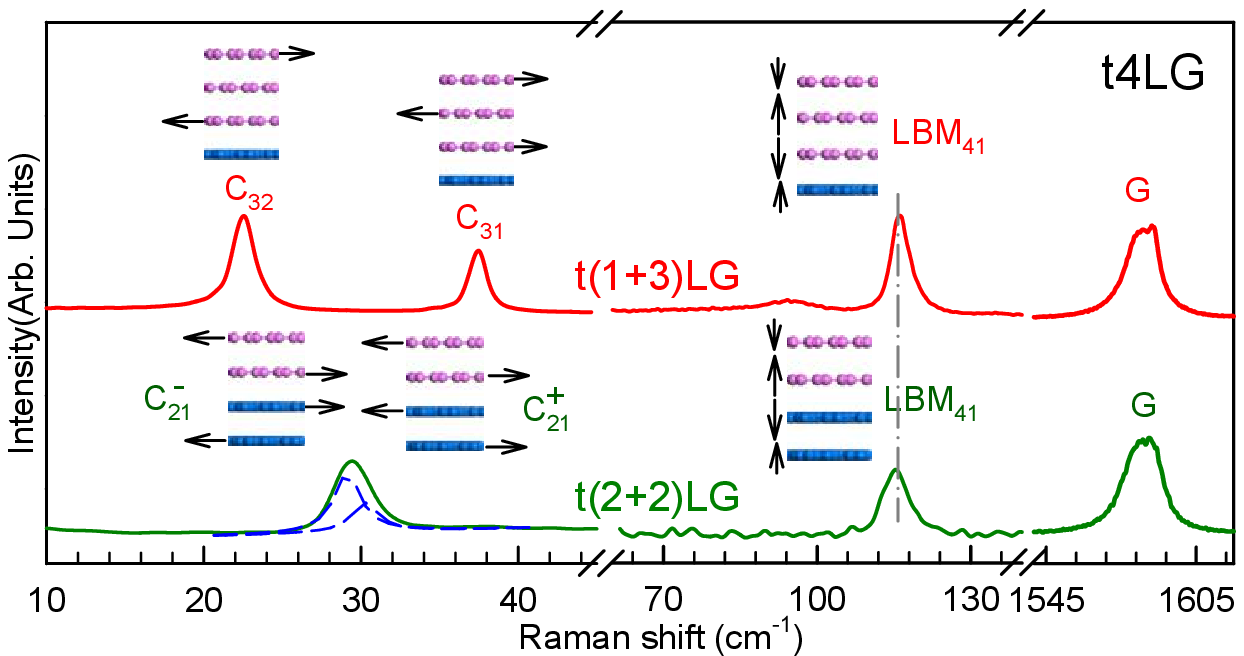}}
\end{figure*}

Layered materials can be assembled to form heterostructures held together by van der Waals interactions. For a given assembly, the relative orientation of the individual layers can change the optical and electronic properties\cite{novops, bonaccorso-2012-production,ferrari-2015-Nanoscale, ponomarenko-2011-tunable,haigh-2012-cross,georgiou-2013-vertical,gong-2014-vertical}. This is also the case when a single material is considered. In multilayer graphene (MLG) samples, for a given number of layers (N), a wide range of properties is accessible by changing the relative orientation of the individual layers\cite{novops, ferrari-2015-Nanoscale, dos-PRL-2007-graphene,trambly-NL-2010-localization,havener-NL-2012-angle,kim-prl-2012-raman,Dresselhaus_jpcl_C13,wu-nc-2014-resonant,cong2014enhanced}. We refer to these as twisted-MLG (tMLG)\cite{wu-nc-2014-resonant}, to indicate a mutual orientation of the planes different from the naturally occurring one\cite{bernal}, with a twist angle ($\theta_t$)\cite{wu-nc-2014-resonant}. The twist vector ($p$,$q$) is defined as the lattice vector of a supercell having $q$, $p$ coordinates with respect to the basis vectors of single layer graphene (SLG)\cite{Laissardiere_NL_2010}. The twist angle can be derived from the twist vector as: $cos{\theta}_{t}$=(${q}^{2}$+$4qp$+${p}^{2}$)/2(${q}^{2}$+$qp$+${p}^{2}$)\cite{Sato-prb-2012,Laissardiere_NL_2010}.

By assembling Bernal stacked\cite{bernal} $m$-layer ($m$LG, $m\geq1$) and $n$-layer ($n$LG, $n\geq1$) flakes, a ($m+n$)-system is formed, which we indicate as t$(m+n)$LG\cite{wu-nc-2014-resonant}. In this notation, a Bernal-stacked BLG is denoted as 2LG, while a twisted one as t(1+1)LG. A flake consisting of a Bernal-stacked BLG placed at a generic angle $\theta_t$ on a Bernal-stacked three layer graphene (TLG) is indicated as t(2+3)LG. This has significantly different properties when compared to a Bernal-stacked 5LG, or to a t(1+4)LG, or t(1+1+3)LG, $etc$, even though all these have the same N=5. For a given total N, the choice of $m$,$n$, $etc.$ (with $m$+$n$+...=N) and relative angles between each m,n,...LGs leads to a family of systems with different optical and electronic properties. Probing the coupling between the interface layers of $m$LG and $n$LG in $t(m+n)$LGs, and its impact on band structure and lattice dynamics, is crucial to gaining fundamental understanding of these systems and to tuning them for novel applications.

Raman spectroscopy is one of the most used characterization techniques in carbon science and technology\cite{acftrans}. The Raman spectrum of graphite and MLG consists of two fundamentally different sets of peaks. Those, such as D, G, 2D, $etc$, present also in SLG, and due to in-plane vibrations\cite{acftrans,tk,FerrariNN}, and others, such as the shear (C) modes\cite{tan-2012-NM-shear} and the layer breathing (LB) modes (LBMs)\cite{Lui1,Sato,FerrariNN}, due to relative motions of the planes themselves, either perpendicular or parallel to their normal. In NLG, all vibrational modes split due to the confinement in the direction perpendicular to the basal plane, $z$, and, for a given N, there are N-1 C or LB modes, which we denote as ${C}_{NN-i}$ and LBM$_{NN-i}$ ($i=1,2,...,N-1$), respectively. Here, ${C}_{N1}$ and LBM$_{N1}$($i.e.$, $i=N-1$) are the C and LB modes with the highest frequencies, respectively. However, due to the low electron phonon coupling (EPC) and different symmetry, it has been not possible, thus far, to detect LBMs for samples at room temperature, unlike the highest energy C modes that can be measured in Bernal-stacked samples at room temperature\cite{tan-2012-NM-shear,lui2014temperature}. In Ref.\citenum{wu-nc-2014-resonant} we have shown that, by performing multi-wavelength Raman spectroscopy on tMLGs, an energy window exists, where a significant intensity enhancement of the C peaks happens, due to resonance with new optically allowed electronic transitions, determined by the relative orientation of the layers. This resonance effect is confirmed by the twist-angle dependence of the G and 2D intensities\cite{havener-NL-2012-angle,kim-prl-2012-raman,Dresselhaus_jpcl_C13}.

Here we directly measure the LBM in tMLGs at room temperature with multi-wavelength Raman spectroscopy, and confirm their assignment by symmetry and polarization analysis combined with density functional theory (DFT). Similar to the C modes, the LBMs exhibit a significant intensity enhancement determined by the relative orientation of the layers. However, unlike the C modes, the observed LBMs are mainly determined by N, which suggests that the breathing coupling at the tMLG interfaces is almost independent of the relative layer orientation. The experimental positions of all LBMs can be described by a linear chain model considering next-nearest interlayer interactions, as verified by DFT. A charge density analysis reveals that the different behavior of C and LB modes in tMLGs is due to the in-plane periodicity mismatch at the twisted interface.
\begin{figure}
\centerline{\includegraphics[width=130mm,clip]{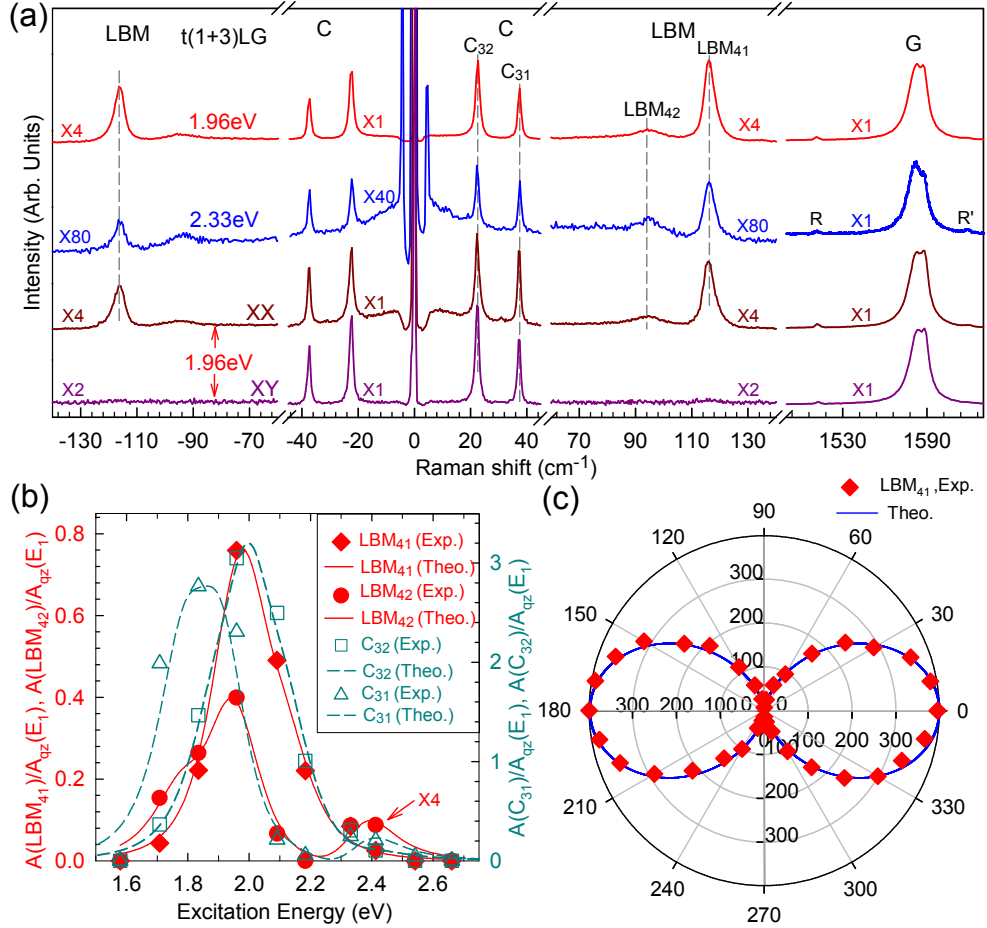}}
\caption{({\bf a}) Stokes/anti-Stokes Raman spectra in the C and LB spectral range, and Stokes Raman spectra in the G peak region for 1.96 and 2.33eV excitation. Polarized Raman spectra are also shown. ({\bf b}) Peak area of $C_{31}$, $C_{32}$ ,LBM$_{41}$ and LBM$_{42}$ as a function of excitation energy. Solid diamonds, open squares and triangles are the experimental data, and solid and dashed lines are the simulations. The peak area of the E$_1$ mode at 127 cm$^{-1}$ of quartz, $A_{qz}(E_1)$, is used to normalize all peaks. ({\bf c})A(LBM$_{41}$) as a function of excitation polarization direction. Open triangles are experimental data and solid lines are the expected trends the symmetry analysis.} \label{Fig1}
\end{figure}
\section{Results and Discussion}
The twisted samples are prepared as follows. Highly oriented pyrolytic graphite (HOPG) is mechanically exfoliated on a Si/SiO$_2$ substrate\cite{novoselov-NAS-2005-two}. During exfoliation mLG flakes are folded onto nLG flakes to form t(m+n)LG flakes, such as the t(1+1+1)LG, t(1+3)LG, t(3+3)LG, t(4+4)LG and t(5+5)LG used in this study. Alternatively, a mLG flake from one substrate can also be transferred onto a nLG flake on another substrate to form t(m+n)LG. Samples t(1+2)LG, t(2+2)LG and t(2+3)LG are prepared in this way. We follow the transfer method described in Ref.\cite{dean-2010-boron}. A flake is exfoliated onto a polymer stack consisting of a water-soluble layer (Mitsubishi Rayon aquaSAVE) and PMMA, and the substrate is floated on the surface of a deionized water bath. During transfer, the target substrate is heated to 110$^{\circ}$C to drive off any water adsorbed on the sample surface, as well as to promote good adhesion of PMMA to the target substrate. N in all initial and twisted MLGs is identified by Raman spectroscopy and optical contrast\cite{Casiraghi-nano-2007,zhao-PRB-2010-charge,ferrari-PRL-2006-raman,wu-nc-2014-resonant}.

Raman spectra are measured in back-scattering at room temperature with a Jobin-Yvon HR800 Raman system, equipped with a liquid-nitrogen-cooled charge-coupled device (CCD), a 100$\times$ objective lens (NA=0.90) and several gratings. The excitation energies are 1.58 and 1.71eV from a Ti:Saphire
laser, 1.96, 2.03, 2.09 and 2.28eV from a He-Ne laser, 1.83, 1.92, 2.18, 2.34 and 2.41eV from a Kr$^+$ laser, and 2.54, 2.67eV from an Ar$^+$ laser. A 1800 lines/mm grating enables us to have each pixel of the charge-coupled detector cover 0.54cm$^{-1}$ at 488nm. Plasma lines are removed from the laser signals using BragGrate Bandpass filters, as described in Ref.\citenum{tan-2012-NM-shear}. Measurements down to 5cm$^{-1}$ for each excitation are enabled by three BragGrate notch filters with optical density 3 and with full width at half maximum (FWHM)=5-10cm$^{-1}$\cite{tan-2012-NM-shear}. The typical laser power is$\sim$0.5mW to avoid sample heating. The accumulation time for each spectrum is$\sim$600s.

We first consider a t(1+3)LG measured at 1.96 and 2.33eV, as for \ref{Fig1}(a). This shows peaks at$\sim$1510 and$\sim$1618cm$^{-1}$. We assign these to the R and R$'$ modes as described in Refs.\citenum{Carozo-peb-2013,Ado2013SSC}. From their position we deduce a ${\theta}_{t}\sim$10.6$^{\circ}$ between the SLG and TLG in this t(1+3)LG, see Methods for details. This corresponds to a twist vector (1,9). Two C modes (${C}_{31}$ and ${C}_{32}$) are observed in t(1+3)LG, mainly localized in 3LG constituent, as previously discussed\cite{wu-nc-2014-resonant}. Two additional modes are observed in t(1+3)LG at$\sim$116 and $\sim$93cm$^{-1}$.

For a given N, the LBM position, Pos(LBM)$_{\textrm{N}}$, can be written as\cite{zhang-PhysRevB-2013,FerrariNN}:
\begin{equation}
\textrm{Pos(LBM)}_{\textrm{N,N-i}}=\textrm{Pos(LBM)}_{\infty} \sin \left[\frac{i\pi}{2\mathrm{N}}\right],
\label{eq:lcm}
\end{equation}
where Pos(LBM)$_{\infty}$ is the LBM in bulk graphite$\sim$128cm$^{-1}$\cite{lattice-graphite-1962}. We note that the N-1 LBM frequencies predicted by \ref{eq:lcm} do not necessarily translate to the experimental observation of the corresponding C and LBM Raman peaks, as these become Raman active under specific selection rules and symmetry constraints, as discussed in Methods.

From \ref{eq:lcm} we get Pos(LBM$_{21}$)=90.5cm$^{-1}$ and Pos(LBM$_{31}$)=110.8cm$^{-1}$. The experimental value 116cm$^{-1}$ is, however, larger than the predicted Pos(LBM$_{31}$), but closer to Pos(LBM$_{41}$)=118cm$^{-1}$. This implies that the LBM is consistent with that of a 4LG, but not with that of the 3LG constituent in the t(1+3)LG, unlike the case of the C modes, where the observed peaks correspond to C$_{31}$ and C$_{32}$\cite{wu-nc-2014-resonant}, as indicated in 1. Thus, we assign the two LBMs in t(1+3)LG as LBM$_{41}$ and LBM$_{42}$. Unlike the D and 2D modes, the LBMs are non-dispersive with excitation energy, $E_{ex}$, as shown in  \ref{Fig1}(a). This is expected, since they come from the Brillouin zone (BZ) center. The peak area of LBM$_{41}$, A(LBM$_{41}$) measured at 1.96eV is$\sim$30 times higher than at 2.33eV, indicative of a resonant Raman behavior. We assign the LBM$_{41}$ and LBM$_{42}$ enhancement to resonance with new optically allowed electronic transitions in t(1+3)LG, as in the case of the C and G modes discussed in Ref.\citenum{wu-nc-2014-resonant}. The C and LB modes are normalized to the E$_1$ mode of quartz\cite{Quartz-nature-1945}. Its position ($\sim$127cm$^{-1}$) is so small that the CCD efficiency difference between C, LB and E$_1$ modes for each excitation energy can be ignored. The resonant profile of LBM$_{41}$ is almost identical to that of C$_{32}$, and the profile of LBM$_{42}$ is similar to that of C$_{31}$, as shown in \ref{Fig1}(b). This indicates that the LBM$_{41}$ resonant behavior can be also assigned to the resonance between the van Hove singularities in the joint density of states of all optically allowed transitions in t(1+3)LG and the laser excitation energy, similar to the C modes in tMLGs\cite{wu-nc-2014-resonant}.

\ref{Fig1}(a) shows that the C and G modes are present in both parallel (XX) and cross (XY) polarization. However, the LBMs in t(1+3)LG vanish in the XY configuration. This can be explained as follows. A t$(m+n)$LG $(m\neq n$) has a ${C}_{3}$ symmetry, and the corresponding irreducible representation\cite{loudon-1964-raman} is $\Gamma$=$A$+$E$. All LBMs have A symmetry, all of C modes have E symmetry, and both the A and E modes are Raman active\cite{loudon-1964-raman}. The A Raman tensor is\cite{loudon-1964-raman}:
\begin{equation}
A=\left[
  \begin{array}{ccc}
    a & 0 & 0\\
    0 & a & 0\\
    0 & 0 & b\\
  \end{array}
\right]
\label{eq:displ}
\end{equation}
This implies that, in backscattering, all LBMs should not be seen in the XY configuration, see Methods, and that their intensity is a function of the angle ($\phi$) between the polarization of the incident light and the polarization (Y) of the Raman signal, $I(LBM)={a}^{2}{cos(\phi)}^{2}$ (see Methods). \ref{Fig1}(c) plots $I$(LBM$_{41}$) as a function of $\phi$. The experimental data (open triangles) are in good agreement with the symmetry analysis.
\begin{figure}
\centerline{\includegraphics[width=140mm,clip]{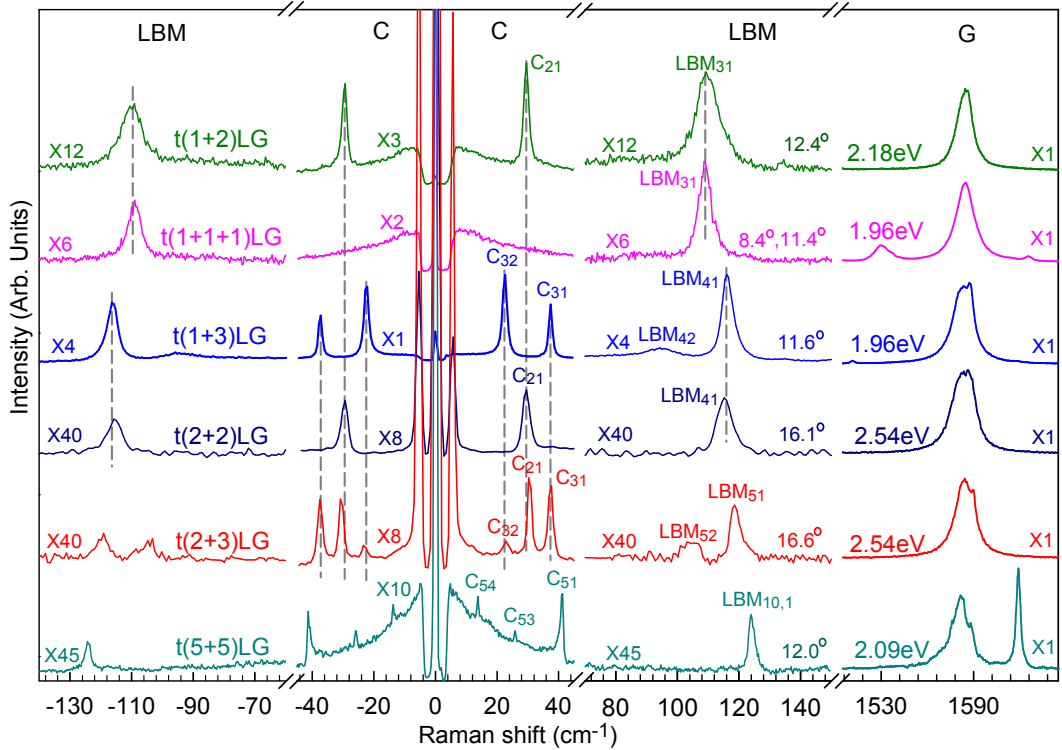}}
\caption{Stokes/anti-Stokes Raman spectra in the C and LB peak region and Stokes spectra in the G spectral region for six tMLGs. $E_{\mathrm{ex}}$ is also indicated. The spectra are scaled and offset for clarity. The scaling factors of the individual spectra are shown. Vertical lines are guides to the eye.} \label{Fig2}
\end{figure}

\ref{Fig2} plots the Raman spectra of six tMLGs: t(1+2)LG, t(1+1+1)LG, t(1+3)LG, t(2+2)LG, t(2+3)LG and t(5+5)LG. To facilitate comparison, all are normalized to have the same intensity of the G peak, I(G). The spectra show the C modes of $m$LG ($m>$1) and $n$LG ($n>$1), localized inside the $m$LG or $n$LG constituents\cite{wu-nc-2014-resonant}. However, this it is not the case for the LBMs. E.g., in t(1+1+1)LG there is no observable C mode, because the twisted interface significantly weakens the shear coupling and pushes the C frequency towards the Rayleigh line, outside the measured spectral region\cite{wu-nc-2014-resonant}. However, in the LBM region, t(1+1+1)LG shows a peak at$\sim$108.8cm$^{-1}$, close to the predicted LBM$_{31}\sim$110.8cm$^{-1}$. A similar peak at $\sim$109.9cm$^{-1}$ is observed in t(1+2)LG. Since both t(1+1+1)LG and t(1+2)LG are two possible t3LG embodiments, we assign the two LBMs in t(1+1+1)LG and t(1+2)LG to LBM$_{31}$. The t(2+2)LG sample shows a LBM$\sim$115.5cm$^{-1}$, very close to the observed$\sim$116cm$^{-1}$ in t(1+3)LG, and to the expected value for LBM$_{41}$. However, unlike t(1+3)LG, t(2+2)LG has a D$_3$ symmetry, and LBM$_{41}$ and LBM$_{43}$ are Raman-active A$_1$ modes, while LBM$_{42}$ is a Raman-inactive A$_2$ mode, see Methods. Thus, LBM$_{42}$ in t(2+2)LG is not detected due to symmetry. In a similar way, we assign the LBMs in t(2+3)LG and t(5+5)LG as LBM$_{51}$, LBM$_{52}$ and LBM$_{10,1}$, respectively. Based on symmetry, all C modes in t($m+n$)LGs are Raman active. Consequently, the C modes of the Bernal-stacked constituents are also observed, such as C$_{51}$, C$_{53}$ and C$_{54}$ in t(5+5)LG.

The above data suggest that, unlike the the C modes, Pos(LBM$_{N,N-i}$) in a t$N$LG ($N=m+n+...$) is mainly determined by N and not by the number of layers of the individual Bernal-stacked constituents ($m$,$n$,...). This means that the LBMs in tMLG are not localized inside its constituents, but are a collective motion involving all layers. We stress that ${\theta}_{t}$ for the six tMLGs in 2 is not the same, as determined by the respective R$'$ and R positions. Various ${\theta}_{t}$ give different band structures with different values for optically-allowed resonance transitions\cite{kim-prl-2012-raman,wu-nc-2014-resonant}. Therefore, for each sample we detect LBMs at different excitations.
\begin{figure}
\centerline{\includegraphics[width=140mm,clip]{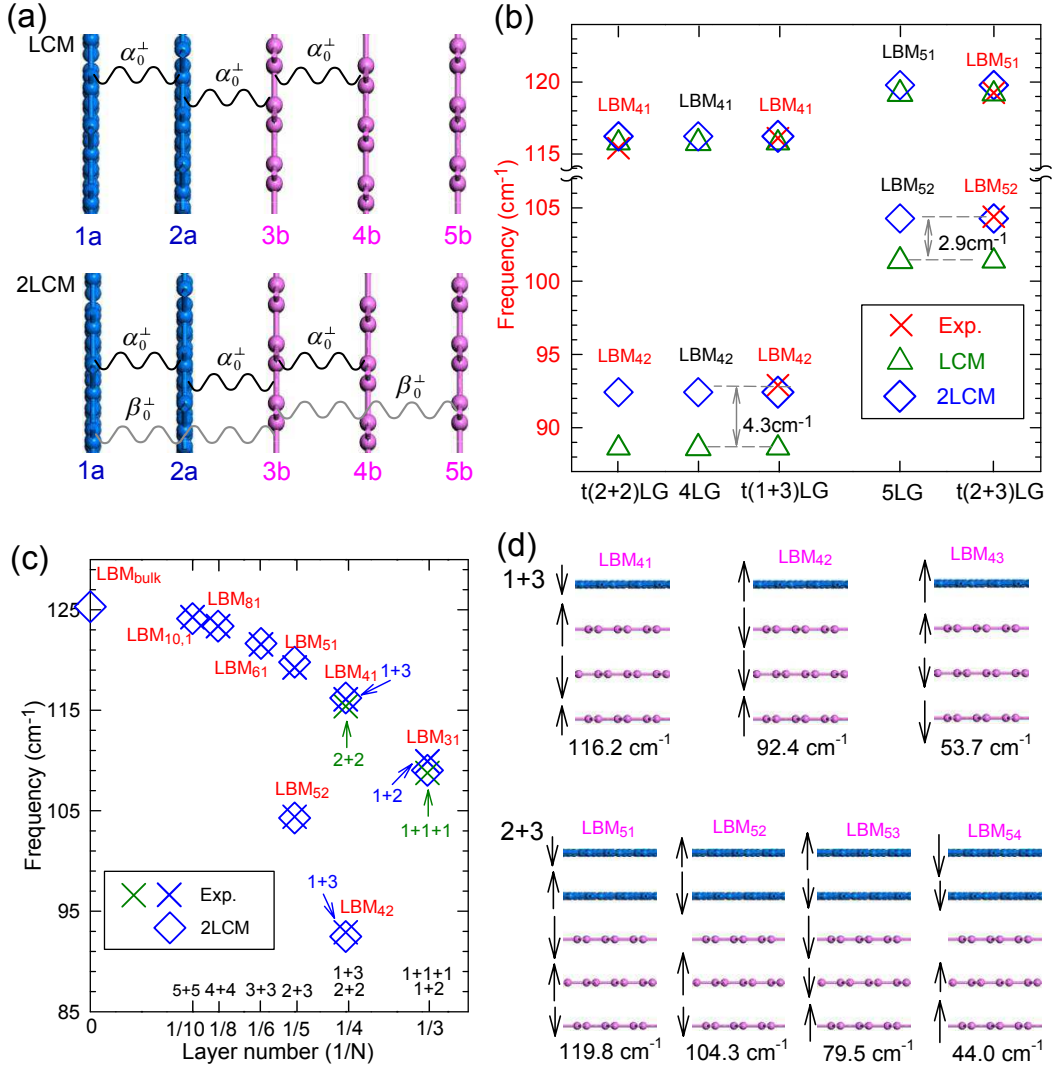}}
\caption{({\bf a})Linear chain model (LCM) and LCM with {\it second-nearest} interlayer coupling (2LCM). ({\bf b}) Theoretical (LCM, open triangles; 2LCM, open diamonds) Pos(LBM$_{N1}$) and Pos(LBM$_{N2}$) in 4LG and 5LG, and experimental (Exp., crosses) and theoretical (2LCM, open diamonds) Pos(LBM$_{N1}$) and Pos(LBM$_{N2}$) in t(2+2)LG, t(1+3)LG and t(2+3)LG. ({\bf c}) Experimental (Exp., open crosses) and theoretical (2LCM, open diamonds) Pos(LBM$_{N1}$) and Pos(LBM$_{N2}$) in tNLG. ({\bf d}) Normal mode displacements and frequencies of t(1+3)LG and t(2+3)LG based on the 2LCM.} \label{Fig3}
\end{figure}

We now consider the effects of changing interlayer interactions on the LBM positions. To do so, we solve the equation of motion for a linear chain system\cite{tan-2012-NM-shear}. The frequencies $\omega$ (in cm$^{-1}$) and displacement patterns can be calculated by solving linear homogeneous equations\cite{tan-2012-NM-shear,wu-nc-2014-resonant}:
\begin{equation}
\omega_{i}^2\mathbf{u}_{i}=\frac{1}{2{\pi }^{2}{c}^{2}\mu }\mathbf{D}\mathbf{u}_{i},
\label{lcm0}
\end{equation}
where $\mathbf{u}_{i}$ is the phonon eigenvector of the $i$th mode with frequency $\omega_i$, $\mu$=7.6$\times$10$^{-27}$kg\AA$^{-2}$ is the SLG mass per unit area, $c$=3.0$\times$10$^{10}$cm s$^{-1}$ is the speed of light, and \textbf{D} is the force constant matrix. In our previous works, we adopted a simple linear chain model (LCM) with only nearest-neighbor interlayer interactions\cite{tan-2012-NM-shear, zhang-PhysRevB-2013}. This allowed us to explain the observed C modes in Bernal and tMLGs, as well as the LBMs in several 2d materials\cite{tan-2012-NM-shear, lui2014temperature, zhang-PhysRevB-2013, zhao-2013-NL-interlayer}. For tMLGs, this also predicts the C modes by introducing a weaker shear force constant (${\alpha}_{t}^{\parallel}$) at the twisted interface\cite{wu-nc-2014-resonant}.

The top panel of  \ref{Fig3}(a) plots the schematic LCM for LBMs in t(2+3)LG if only the nearest-neighbor layer-breathing interlayer interaction ($\alpha_0^\perp$) is considered. The experimental frequencies of t(2+2)LG, t(1+3)LG and t(2+3)LG are plotted in  \ref{Fig3}(b) as crosses, and those of all tMLGs are summarized in  \ref{Fig3}(c), including LBMs from t(3+3)LG and t(4+4)LG, whose Raman spectra are presented in Methods. By taking the average frequency (115.8cm$^{-1}$) of the experimental LBM$_{41}$ measured in t(1+3)LG and t(2+2)LG, we get $\alpha_{0}^{\perp}$=106$\times{10}^{18}$ Nm$^{-3}$, which would give 119.2cm$^{-1}$ for Pos(LBM$_{51}$) in 5LG, consistent with the value measured in t(2+3)LG.  \ref{Fig3}(b) also gives Pos(LBM$_{42}$)=88.6cm$^{-1}$ for 4LG and Pos(LBM$_{52}$)=101.4cm$^{-1}$ for 5LG, which are 4.3 and 2.9cm$^{-1}$ lower than those observed in t(1+3)LG and t(2+3)LG, respectively.

These lower frequencies suggest that the LCM, with only nearest-neighbor interlayer interactions, may be insufficient to reproduce the interlayer breathing coupling in tMLGs. If a weakened coupling at the twisted interface is included in the LCM, it will result in LBM red-shift for both LBM$_{N1}$ and LBM$_{N2}$ ($N$=4,5), making the agreement worse, see Methods. We thus introduce an interlayer breathing force constant between the second-nearest neighbor layers (${\beta}_{0}^{\perp}$). The new model is denoted as 2LCM, and is schematically shown in  \ref{Fig3}(a) for LBMs in t(2+3)LG. 2LCM with ${\beta}_{0}^{\perp}\sim9.3\times{10}^{18}$ Nm$^{-3}$ fits the experimental data best, as indicated by diamonds in  \ref{Fig3}(b). With 2LCM we can well fit the frequencies of the observed LBMs in all tMLGs, as shown in  \ref{Fig3}(b,c). Additionally, we can expand the 2LCM predictions to bulk graphite, based on the parameters fitted on our experiments. The silent LBM (${B}_{2g}$) in graphite is derived to be $\sim$125.3cm$^{-1}$, very close to $\sim$128cm$^{-1}$ determined by neutron spectrometry\cite{lattice-graphite-1962}.

The normal mode displacements and frequencies of each LBM in t(1+3)LG and t(2+3)LG as derived by the 2LCM are summarized in  \ref{Fig3}(d). In LBM$_{N,1}$, the relative motions of the nearest-neighbor layers are always out-of-phase, and those of the second-nearest-neighbor layers are always in-phase. This would suggest Pos(LBM$_{N,1}$) to be insensitive to the second-nearest-neighbor interlayer coupling. However, the relative motions of the second-nearest-neighbor are out-of-phase for LBM$_{42}$ in t(1+3)LG and LBM$_{52}$ in t(2+3)LG. Thus, the reason why \ref{eq:lcm} fits Pos(LBM$_{N1}$) well, but predicts lower frequencies for Pos(LBM$_{N2}$) is, most likely, due to the lack of interaction from second-nearest-neighbor layers.

The 2LCM gives the same LB coupling for twisted and Bernal-stacked interfaces. However, the shear coupling at twisted interfaces is$\sim$20\% of that at Bernal-stacked interfaces\cite{wu-nc-2014-resonant}. We now use DFT and density functional perturbation theory (DFPT)\cite{baroni2001phonons} to validate this model, and to understand the difference between the C and LBMs in tMLG. Because a t$(m+n)$LG with a twist vector of (1,2), $i.e.$, a twist angle of 21.8$^\circ$, is a simplest twist structure, we consider t(2+3)LG and t(1+2)LG with this twist angle for DFPT.
\begin{table}
\caption{{\it Ab initio} interlayer force constants between each couple of layers along $z$ for t(2+3)LG. Twisting happens between the second (denoted 2a) and third (denoted 3b) layers. Two categories of Bernal-stacked layers are grouped as "a" and "b", respectively.}
\begin{center}
\begin{tabular*}{8.2cm}{cccccc}
  \hline\hline
\parbox{2.5cm}{Force constant ($\times{10}^{18}$ Nm$^{-3}$)} & \parbox{1.0cm}{1a} & \parbox{1.00cm}{2a} & \parbox{1.0cm}{3b} & \parbox{1.0cm}{4b} & \parbox{1.0cm}{5b}\\
\hline
1a  & ~~ - & - & - & - & - ~~\\
2a  & ~~114.2 & - & - & - & - ~~\\
3b  & ~~4.3 & 114.7 & - & - & - ~~ \\
4b  & ~~3.9 & 3.4 & 120.1 & - & - ~~ \\
5b  & ~~4.0 & 6.1 & 2.6 & 113.4 & - ~~ \\
\hline\hline
\end{tabular*}
\end{center}
\label{tab:fc}
\end{table}

We first calculate the frequencies of LBMs in t(2+3)LG with a (1,2) twist vector. They are 126.3cm$^{-1}$ (LBM$_{51}$), 107.3cm$^{-1}$ (LBM$_{52}$), 79.9cm$^{-1}$ (LBM$_{53}$) and 47.9cm$^{-1}$ (LBM$_{54}$), respectively, overall consistent, but a few cm$^{-1}$ larger, than the experiments reported in \ref{Fig3}, owing to the slightly overestimated interlayer interaction\cite{thiocu,bn-qmc}. A full comparison between calculated and measured frequencies is reported in Methods. \ref{Fig3}(a) shows the five layers and four interfaces in t(2+3)LG. We denote them as 1a, 2a, 3b, 4b, and 5b from left to right. Twisting happens between layers 2a and 3b, and we call this interface 2a-3b. The interlayer force constant (IFC) along $z$ is a measure of the interlayer breathing coupling and is calculated as for Methods. The IFC along $z$ between 2a and 3b (the twisted interface) is 114.7$\times{10}^{18}$Nm$^{-3}$, close to that of other Bernal stacked interfaces, pointing to a similar breathing coupling at the twisted interface as that of the Bernal stacked interface. We get ${\alpha}_{0}^{\perp}$=115.6$\times{10}^{18}$Nm$^{-3}$ by averaging the computed IFCs along $z$ at the four interfaces, which agrees with the experiments and with the value derived using the 2LCM (116$\times{10}^{18}$Nm$^{-3}$)
\begin{figure}
\centerline{\includegraphics[width=140mm,clip]{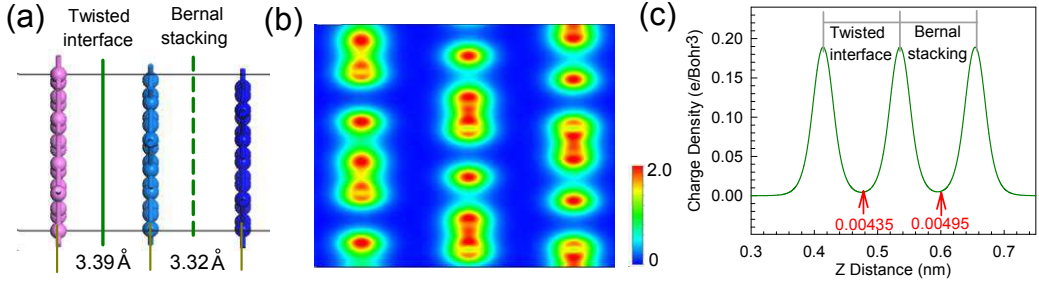}}
\caption{({\bf a}) Atomic structure of t(1+2)LG with $\theta_t$=21.8$^{\circ}$. The twisted and Bernal-stacked interfaces are indicated by green solid and dashed lines, respectively. ({\bf b}) Charge density contour for the t(1+2)LG sample in {\bf a}. ({\bf c}) Layer-averaged charge density along $z$.}
\label{Fig4}
\end{figure}

We now address the substantial force constant difference for the C and LBMs in twisted and Bernal-stacked layers. Van der Waals forces, specifically the dispersion force\cite{dion2004van}, rule the interlayer interactions, and play a key role in the difference between C and LBMs in tMLG. \ref{Fig4}(a) plots the sideview of the fully-relaxed atomic structure of a t(1+2)LG with a (1,2) twist vector. We also consider t(1+2)LGs with twist angles of 13.2$^{\circ}$, 38.2$^{\circ}$ and 46.8$^{\circ}$. Four stacking configuration are considered for each angle. The average interlayer distance for every configuration is 3.39\AA, with a variation less than 0.01\AA. Ref.\cite{Liu-nc-2014-twsited} reported a similar result for twisted MoS$_2$ bilayers, with the calculated interlayer distances nearly identical in the 0$^{\circ}$ to 60$^{\circ}$ range. Our calculations are also consistent with the interlayer distance in t(1+1)LGs calculated in Ref.\cite{prb_twist_angle}, showing a larger interlayer distance at the twisted interface when compared to Bernal-stacked layers, and little correlation between interlayer distance and twist angle. The interlayer distance between the twisted interface and the Bernal-stacked interface in t(1+2)LG is$\sim$0.1\AA~, much smaller than in MoS$_2$/MoSe$_2$ heterostructures ($\sim$0.6\AA)\cite{Kang-NL-MoS2/MoSe2}, where the interface has$\sim$4\% lattice mismatch. This is directly relevant for the out-of-plane breathing vibration along $z$, as represented by the LBM frequency. \ref{eq:disp4} and \ref{eq:disp5} in Methods indicate that the interaction strength has a positive correlation with charge density, nearly identical at the two interfaces of \ref{Fig4}(b). A small difference is revealed by calculating the mean charge densities at the two interfaces. The interlayer breathing interaction at the twisted interface is very close to that of Bernal-stacked interfaces, again supporting the 2LCM.
\begin{figure}
\centerline{\includegraphics[width=140mm,clip]{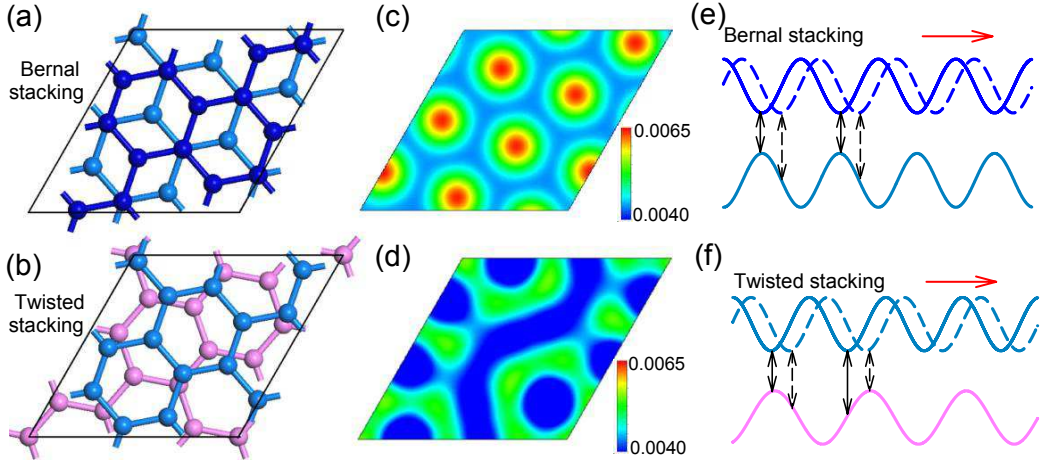}}
\caption{Atomic structure of ({\bf a}) Bernal-stacked 2LG on the top of t(1+2)LG and ({\bf b}) the t2LG at the bottom of t(1+2)LG. The corresponding charge density ({\bf c}) at the Bernal-stacked interface in ({\bf a}) and ({\bf d}) at the twisted interface in ({\bf b}). Schematic diagram for the charge distribution ({\bf e}) at the Bernal-stacked interface in ({\bf a}) and ({\bf f}) at the twisted interface in ({\bf b}). The latter shows the mismatched periodicity between the two layers.}
\label{Fig5}
\end{figure}

We turn to consider the C modes in t(1+2)LG with a (1,2) twist vector. Top views of the Bernal-stacked and twisted interfaces are shown in \ref{Fig5}(a,b), while their corresponding charge densities in the middle of two SLGs is shown in \ref{Fig5}(c,d). Both plots indicate that the C$_{6}$ symmetry at the Bernal-stacked interface is broken at the twisted interface (\ref{Fig5}(b)), and the local density periodicity is also lifted (\ref{Fig5}(d)). Twisting forms a Moir\'e pattern, resulting in a locally mismatched periodicity of the charge density variations. \ref{Fig5}(e,f) plots a schematic diagram illustrating the effect of periodicity mismatch on the C vibrations. In Bernal-stacked interfaces the interatomic restoring forces are all along the positive direction for a small displacement, \ref{Fig5}(e). With the elimination of the local periodicity, a Moir\'e pattern at the twisted interface makes the interatomic restoring forces negative or positive, as shown in \ref{Fig5}(f). Therefore, shear restoring forces are nearly canceled at the twisted interface, resulting in a much weaker shear coupling than in Bernal-stacked interfaces. Thus, the softening of the C modes is due to the periodicity mismatch at the twisted interface.
\section{Conclusions}
We measured by resonant Raman spectroscopy the LBMs of tMLG, an archetype heterostructure. We showed that a second-nearest neighbor linear chain model explains all the measured spectra, as validated by ab-initio calculations. The interlayer shear coupling strength declines at twisted interfaces due to the periodicity mismatch between two twisted layers, while the interlayer breathing coupling remains nearly constant. Beyond tMLGs, the interlayer interaction of other heterostructures can also be measured by Raman spectroscopy\cite{Zhang_2015_csr}. Unlike graphene, the interlayer coupling modes of other 2d layered materials, like transition metal chalcogenides\cite{zhang-PhysRevB-2013,zhao-2013-NL-interlayer,Zhang_2015_csr,Anharmonic-mos2-prb} ($e.g.$ MoS$_{2}$ and WSe$_{2}$) and others, such as NbSe$_{2}$\cite{yokoya_2001_fermi}, and Bi$_{2}$Se$_{3}$\cite{Zhang_2015_csr} and Bi$_{2}$Te$_{3}$\cite{Zhang_2015_csr}, can be measured more easily, due to the stronger electron-phonon coupling. Therefore, the LBMs should be also measurable in heterostructures with clean interfaces, such as graphene/MoS$_{2}$, graphene/WS$_{2}$, MoS$_{2}$/WSe$_{2}$, thus allowing one to probe the interlayer coupling of these two-component layered heterostructures and, possibly, even more complex structures. By studying both C and LB modes together, it should be possible to detect the detailed components, number of layers of each component, and the coupling amongst the components, a crucial step for both fundamental science and technology based on these materials.

\begin{acknowledgement}
We acknowledge support from the special funds for Major State Basic Research of China, contract No. 2012CB932704, the National Natural Science Foundation of China, grants 11225421, 11474277 and 11434010, 11274380 and 91433103, the Program for New Century Excellent Talents in University, the Physics Lab of High-Performance Computing of Renmin University of China and Shanghai Supercomputer Center, the EU Graphene Flagship (no. 604391), ERC Grant Hetero2D, EPSRC Grants EP/K01711X/1, EP/K017144/1, EU grant GENIUS, a Royal Society Wolfson Research Merit Award
\end{acknowledgement}
\section{Methods}
\subsection{Calculations}
Structural relaxation and charge density calculations are performed using the DFT code Vienna ab-initio simulation package (VASP)\cite{kresse1996efficient} within the projector augmented wave method\cite{blochl1994projector,kresse1999ultrasoft} and a plane-wave basis. The exchange-correlation potential is treated within the generalized gradient approximation. Van der Waals interactions are considered under the framework of the vdW-DF method\cite{dion2004van} with the optB86b exchange functional\cite{klimevs2011van}. This exchange-correlation combination is more accurate in predicting lattice parameters in 2d materials, such as black phosphorous\cite{bp2014} and boron nitride\cite{bn-qmc} than other vdW-DF approaches, while it is known to slightly overestimate interlayer binding energy\cite{bn-qmc,thiocu}. In vdW-DF the description of the dispersion force requires the inclusion of the non-local correlation energy\cite{dion2004van}:
\begin{equation}
{E}^{nl}_{c}=\frac{\hbar}{2}\int\int{\rm d}\mathbf{r}{\rm d}\mathbf{r'}n(\mathbf{r})\Phi(\mathbf{r},\mathbf{r'})n(\mathbf{r'})
\label{eq:disp4}
\end{equation}
\begin{equation}
\Phi(\mathbf{r},\mathbf{r'})\rightarrow \frac{3{e}^{4}}{2{m}^{2}{\omega}_{0}(\mathbf{r}){\omega}_{0}(\mathbf{r'})[{\omega}_{0}(\mathbf{r})+{\omega}_{0}(\mathbf{r'})]{d}^{6}}
\label{eq:disp5}
\end{equation}
with $n(r)$ the charge density, $\Phi$ the correlation interaction kernel and $d$ the distance between two SLGs. For $d\rightarrow\infty$, $\Phi\propto{n}^{-1.5}{d}^{-6}$, which means ${E}^{nl}_{c}\propto{n}^{0.5}{d}^{-6}$. The non-local correlation energy between two SLGs is determined by charge density and layer distance.

A 29$\times$29$\times$1 k-mesh is used to sample the BZ for Bernal-stacked supercells and an 11$\times$11$\times$1 one for twisted supercells, due to the $\sqrt{7}$ larger lattice constant. The energy cutoff for the plane-wave basis is 400eV. All atoms are fully relaxed until the residual force per atom is smaller than 0.001eV$\cdot$\AA$^{-1}$. Vibrational frequencies are calculated using DFPT\cite{baroni2001phonons}, as implemented in VASP. In an interlayer vibrational mode, the whole layer can be treated as one rigid body. The IFC is constructed by summing inter-atomic force constants over all atoms from each of the two adjacent layers. The matrix of inter-atomic force constants, essentially the Hessian matrix of the Born-Oppenheimer energy surface, is defined as the energetic response to a distortion of atomic geometry in DFPT\cite{baroni2001phonons}.

\begin{figure}
\centerline{\includegraphics[width=80mm,clip]{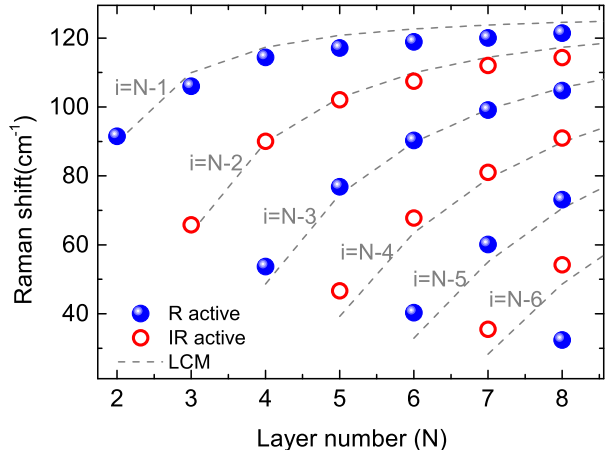}}
\caption{LBMs for Bernal stacked NLGs. Gray lines indicate the calculated LCM.}\label{Fig6}
\end{figure}

\ref{Fig6} plots the LBM positions calculated from the LCM in \ref{eq:lcm} and by DFT for Bernal Stacked samples (with DFT data rigidly shifted by $\sim$10cm$^{-1}$) implemented in the QuantumESPRESSO package\cite{QE}. The in-plane lattice constant is set to 2.43~\AA{} and the interlayer distance 3.26~\AA{} to match the experimental ZO$'$ frequency at the $\Gamma$ point. A norm-conserving Martins-Troullier pseudopotential within the local density approximation (LDA) is used, and the plane waves were expanded up to a 80Ry cutoff. The BZ is sampled using a 12$\times$12$\times$4 Monkhorst-Pack mesh and Methfessel-Paxton smearing with 0.03Ry width is used for the electronic occupations close to the Fermi level. The dynamical matrices are computed on a 8$\times$8$\times$3 mesh. The modes are either Raman (R) or Infrared (IR) active.

\begin{figure}
\centerline{\includegraphics[width=80mm,clip]{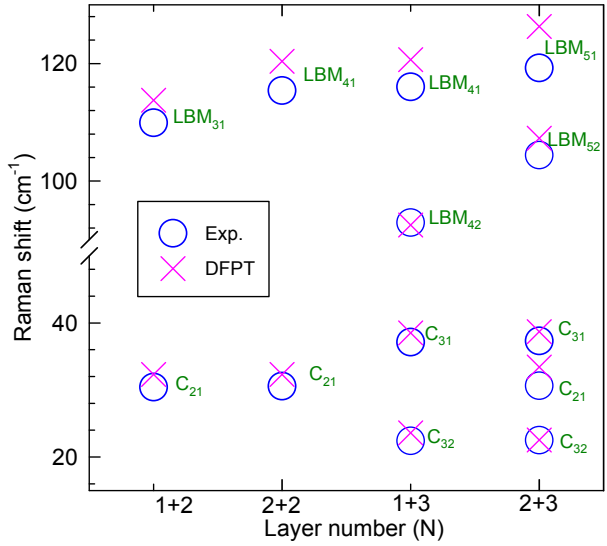}}
\caption{C and LBMs in tMLGs. (pink crosses) DFPT data. (blue circles) Experiments.} \label{Fig7}
\end{figure}

\ref{Fig7} compares DFPT and experimental Pos(C) and Pos(LBM) in various tMLGs.

\begin{figure}
\centerline{\includegraphics[width=100mm,clip]{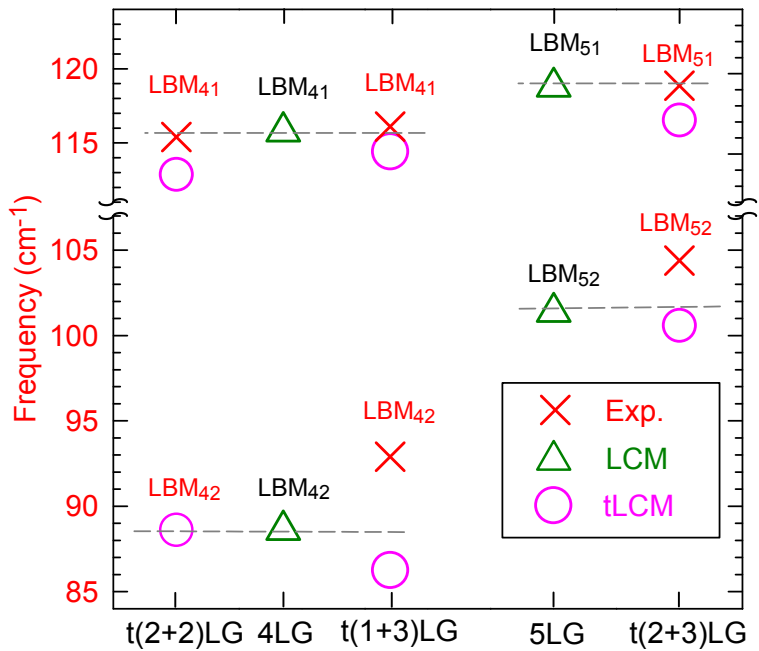}}
\caption{(Crosses) Experimental data. (Up triangles) LCM. (Circles) tLCM.}\label{Fig8}
\end{figure}

\ref{Fig8} compares the experimental LBMs in tMLGs with those calculated with the LCM of \ref{eq:lcm} and those using a LCM with a weakened coupling at the twisted interface (tLCM). A 10\% weakened coupling red-shifts both LBM$_{N1}$ and LBM$_{N2}$ (N=4 and 5), resulting in a worse fit to the experimental data.
\subsection{Relation between $\theta_t$ and the frequency of R and R$'$ modes}
The observation of R and R$'$ peaks in the Raman spectra of tBLG is due to the superlattice modulation activating phonons in the BZ interior\cite{Carozo-peb-2013,Ado2013SSC}. $\theta_t$ dictates the wavevector for this modulation, with $q(\theta_t)$ the difference between the basic vectors of two SLGs in the BZ. The wavevector $q(\theta_t)$ selects the phonons along the phonon dispersion that become Raman active. The relation between ${q}_{\Gamma K}$$(\theta_t)$ and the $\theta_t$ is given by:
\begin{equation}
{q}_{\Gamma K}(\theta_t)=\frac{4\pi }{3a}\left(1-\sqrt{7-2\sqrt{3}sin{\theta}_{t}-6cos{\theta}_{t}} \right),
\label{eq:disp6}
\end{equation}
where $a$=2.46\AA~is the SLG lattice constant. From Pos(R) and Pos(R$'$), ${q}_{\Gamma K}$$(\theta_t)$ can be determined from the SLG phonon dispersion. \ref{eq:disp6} then gives $\theta_t$. For the assignment, we use phonon dispersions calculated from DFT\cite{piscanec} corrected with GW (Green$'$s function G of the screened Coulomb interaction W), which well reproduces the experimental LO-TO splitting\cite{PhysRevB.80.085423,PhysRevB.78.081406}.

\begin{figure}
\centerline{\includegraphics[width=100mm,clip]{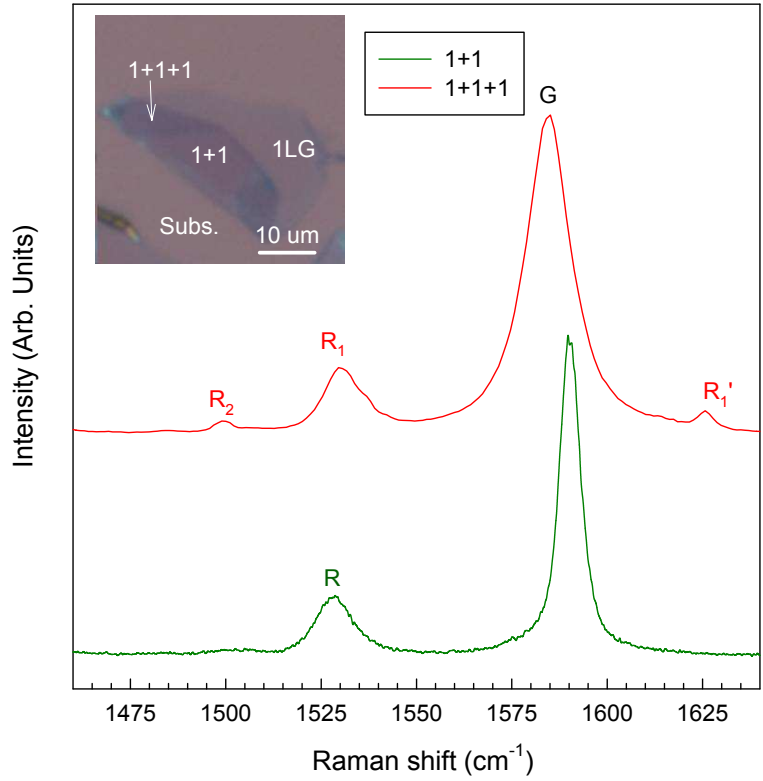}}
\caption{Optical image and Raman spectra of t(1+1+1)LG and t(1+1)LG.} \label{Fig9}
\end{figure}

\ref{Fig9} plots the optical image and Raman spectra of t(1+1+1)LG and t(1+1)LG. There are two couples of R and R$'$ modes in t(1+1+1)LG due to twice folding a SLG. The R$_{1}$ mode of t(1+1+1)LG is at 1529cm$^{-1}$, the same position as the R mode of t(1+1)LG. This means that the R$_{1}$ and R$_{1}$$'$ are from the bottom twisted bilayer of t(1+1+1)LG and that R$_{2}$ is from the top twisted bilayer of t(1+1+1)LG.

\begin{figure}
\centerline{\includegraphics[width=120mm,clip]{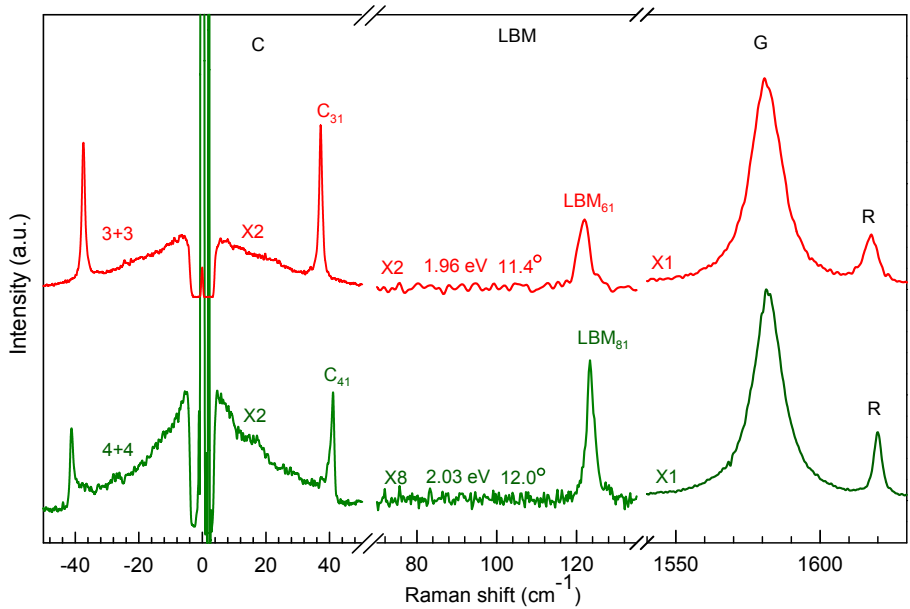}}
\caption{Stokes/anti-Stokes Raman spectra in the C peak region and Stokes spectra in the LBM and G spectral regions for t(3+3)LG and t(4+4)LG. The twist angle and laser energy is marked for each sample.} \label{Fig10}
\end{figure}

\ref{Fig10} plots the optical image and Raman spectra of t(3+3)LG and t(4+4)LG. $\theta_t$ of t(3+3)LG and t(4+4)LG are 11.4$^\circ$ and 12.0$^\circ$, respectively, determined by the respective R$'$ modes.
\subsection{Symmetry and Raman activity of C and LBMs in t$(m+n)$LG $(m\neq n$) and t$(n+n)$LG $(n\geq 2$)}
t$(m+n)$LG $(m\neq n$) have C$_{3}$ symmetry, the corresponding irreducible representation is $\Gamma$=$A$+$E$, and both $A$ and $E$ modes are Raman active\cite{loudon-1964-raman}. In t$(m+n)$LG with $(m\neq n$), all non-degenerate LBMs have $A$ symmetry, and all of double-degenerate C modes belong to $E$ symmetry\cite{loudon-1964-raman}.

t$(n+n)$LG $(n\geq 2$) have D$_{3}$ symmetry, and the corresponding irreducible representation is $\Gamma$=${A}_{1}$+${A}_{2}$+$E$\cite{loudon-1964-raman}. ${A}_{1}$ and $E$ modes are Raman active, while ${A}_{2}$ are Raman inactive\cite{loudon-1964-raman}. In t(2+2)LG, LBM$_{41}$ and LBM$_{43}$ have ${A}_{1}$ symmetry, while LBM$_{42}$ has ${A}_{2}$ symmetry, and all the C modes are $E$\cite{loudon-1964-raman}.

The Raman intensity is proportional to $|$${e}_{i}\cdot {R}_{t} \cdot{e}_{s}$$|$$^{2}$, where ${e}_{i}$ and ${e}_{s}$ are the unit vectors describing the polarizations of the incident and scattered light, and ${R}_{t}$ is Raman tensor\cite{loudon-1964-raman}. In our work, the polarization of the incident light is at an angle ($\phi$) set by a $\lambda$/2 wave plate (${e}_{i}$=[cos$\phi$ sin$\phi$ 0]), and the polarization of the scattered light is fixed along the horizontal (${e}_{s}$=[1 0 0]$'$). Therefore, the Raman tensors of for the LBMs in t$(m+n)$LG $(m\neq n$) is\cite{loudon-1964-raman}:
\begin{equation}
A=\left[
  \begin{array}{ccc}
    a & 0 & 0\\
    0 & a & 0\\
    0 & 0 & b\\
  \end{array}
\right]
\label{eq:displ1}
\end{equation}
Thus, I(LBMs) in t$(m+n)$LG $(m\neq n$) is:
\begin{equation}
I(LBMs)\propto {\begin{vmatrix}[cos\phi~sin\phi~0]\begin{bmatrix}
a & 0 & 0 \\
0 & a & 0 \\
0 & 0 & b
\end{bmatrix} \begin{bmatrix}
1\\
0\\
0
\end{bmatrix} \end{vmatrix}}^{2}={a}^{2}{cos(\phi)}^{2}
\label{eq:disp44}
\end{equation}
As discussed above, the LBMs in the t(1+3)LG are Raman active, except LBM$_{42}$ (Raman inactive). LBM$_{41}$ and LBM$_{43}$ in t(2+2)LG are Raman active. Both LBM$_{41}$  and LBM$_{42}$ are observed in t(1+3)LG, see \ref{Fig2}. However, only LBM$_{41}$ is observed in t(2+2)LG. The absence of LBM$_{43}$ in t(1+3)LG and t(2+2)LG may result from a weaker EPC\cite{tan-2012-NM-shear}. The Raman tensor of the ${A}_{2}$ mode in t(2+2)LG is the same as that of the $A$ mode in t(1+3)LG\cite{loudon-1964-raman}, thus the I(LBM$_{41}$) in t(2+2)LG is also laser-polarization dependent.

\section{Corresponding Author}
E-mail address: wji@ruc.edu.cn,~phtan@semi.ac.cn

\section{Author Contributions}
P.-H.T. conceived the research. W.-P.H., Y. L., W. S. and X.-F.Q. prepared the samples. J.-B.W. and
P.-H.T. performed Raman measurements. J.-B.W., Z.-X.H., X. Z., M. I., S. M., A.C.F., W.J. and P.-H.T. performed the theoretical analysis. P.-H.T., A.C.F., J.-B.W., W.J. and Z.-X.H. wrote the manuscript, which all authors read and commented on.

\section{Notes}
The authors declare no competing financial interests.

\providecommand*\mcitethebibliography{\thebibliography}
\csname @ifundefined\endcsname{endmcitethebibliography}
  {\let\endmcitethebibliography\endthebibliography}{}

\end{document}